# *Identifying Critical LMS Features for Predicting At-risk Students*

Ying Guo, Cengiz Gunay, Sairam Tangirala, David Kerven, Wei Jin, Jamye Curry Savage and Seungjin Lee

Abstract

Learning management systems (LMSs) have become essential in higher education and play an important role in helping educational institutions to promote student success. Traditionally, LMSs have been used by postsecondary institutions in administration, reporting, and delivery of educational content. In this paper, we present an additional use of LMS by using its data logs to perform data-analytics and identify academically at-risk students. The data-driven insights would allow educational institutions and educators to develop and implement pedagogical interventions targeting academically at-risk students. We used anonymized data logs created by Brightspace LMS during fall 2019, spring 2020, and fall 2020 semesters at our college. Supervised machine learning algorithms were used to predict the final course performance of students, and several algorithms were found to perform well with accuracy above 90%. SHAP value method was used to assess the relative importance of features used in the predictive models. Unsupervised learning was also used to group students into different clusters based on the similarities in their interaction/involvement with LMS. In both of supervised and unsupervised learning, we identified two most-important features (Number_Of_Assignment_Submissions and Content_Completed). More importantly, our study lays a foundation and provides a framework for developing a real-time data analytics metric that may be incorporated into a LMS.

## 1 INTRODUCTION

A Learning Management System (LMS) refers to a web-based information system that facilitates access to learning content and that is managed by educational institutions' administration [5]. LMSs exemplify a paradigm shift in the ways that pedagogical challenges are addressed through technological solutions. Since the introduction of web-accessible LMSs, they have undergone very rapid adoption in educational institutions throughout the world. According to a previous report [20], the global LMS market size was around 10.84 billion USD in 2020. This demand in the use of LMS systems experienced a tremendous spike in 2020, due to the global impact of COVID-19. It was also reported that its global market exhibited a growth of 23.8% in year 2020 as compared to the average yearly growths during 2017-19. Furthermore, the consumer market for LMS is projected to grow from USD 13.38 billion in 2021 to USD 44.49 billion in 2028, after accounting for the returning of global economies to pre-pandemic levels after COVID-19 period [20].

In today's high-tech economy which is becoming increasingly dependent on insights generated from data, it is especially important for educational institutions to leverage the value that can be obtained from the vast volumes of data that is systematically logged and collected from their deployed LMSs. In a February 2020 survey conducted jointly by Jenzabar Inc. (a technology innovator in higher education domain) and University Business® (education-news content provider) [21], a total of 175 higher education leaders from a variety of sizes and types of institutions in the U.S. were surveyed. The demographic distribution of the functional titles of the respondents included President/Vice President

(21%), followed by Dean/Faculty (16%), Provost/Vice Provost (11%), Enrollment/Admissions/Registrar (11%), Finance/Business Officer (8%), and Institutional Research (7%). In this survey, it was reported that, approximately 80% of higher education institutions' goals were to improve their student success and academic-program completion rates. Another 74% of institutions reported to have student-enrollment as their strategic priority. We believe that LMS data is extremely valuable to every educational institution, and it may be used in designing strategic plans for addressing and achieving the respective institution's needs and goals.

While serving as an irreplaceable pedagogical resource, a traditionally limited use of LMS system (for managing pedagogy content, recording grades, automating class-related tasks, etc.) may not fully support educational institution's strategic goals in terms of student's academic progress, retention rate, and improving overall teaching and learning experience. In this article, we present insights generated from the LMS logged data used by our college and propose ways of using data science as a tool to promote educational institutions goals of student success. We believe that the knowledge obtained from analyzing and carefully studying LMS logged data can be effectively used by an educational institution to offer an improved academic experience to its students.

### 1.1  About Our College (OC) [anonymized]

OC is a public, four-year open-access institution, located in [anonymized]. OC is the newest institution in the [anonymized]. It opened its doors in fall, 2006 with fifteen faculty members, a student headcount of 118, and three majors [2]. Since its inception, OC has quickly expanded in the areas of faculty count, student enrollment, and number offered programs and evolved into a campus awarding degrees in nineteen majors across five academic schools with a student body about one hundred times the size of its initial fall 2006 class.

[anonymized] is the second most populous county in [anonymized] and is the seventh most ethnically diverse county [19] in the [anonymized] according to 2020 census data [9]. OC's student body reflects a similar ethnic diversity. As shown in Table 1, for the years 2018 - 2020, OC has been a minority majority institution as demonstrated by the corresponding fall semester demographic information [3, 4].

Table 1: Fall Student Race/Ethnicity Demographics

| Year (Fall) | American Indian/ Alaskan | Asian | Black/African American | Hispanic / Latino | Multi-Racial | Native Hawaiian/ Pacific Islander | White | Unknown |
|---|---|---|---|---|---|---|---|---|
| 2018 | <1% (13) | 11% | 32% | 21% | 4% | <1% (28) | 31% | 1% |
| 2019 | <1% (17) | 11% | 32% | 24% | 4% | <1% (21) | 28% | 1% |
| 2020 | <1% (16) | 11% | 33% | 25% | 4% | <1% (23) | 27% | <1% (80) |

As shown in Table 2, about two-thirds of the student body took full-time course load during the fall semester of 2018 - 2020. The student headcounts and breakdown by full-time and part-time faculty percentages are also shown in Table 2 [3, 4].

Table 2: Fall Student Headcounts and Full-time/Part-time Status

| Year (Fall) | Student Headcount | Full-Time Faculty | Part-Time Faculty |
|---|---|---|---|
| 2018 | 12,508 | 66% | 34% |
| 2019 | 12,831 | 67% | 33% |
| 2020 | 11,627 | 65.5% | 34.5% |

During the fall semesters of 2018 - 2020, the majority of students fell into the 18 - 22 year-old age range, as shown in Table 3. The age distribution of OC's student population for the same period is also shown in Table 3 [3, 4].

Table 3: Fall student enrollment by age

| Year (Fall) | Under 18 | 18-22 | 23-34 | 35-44 | Over 44 |
|---|---|---|---|---|---|
| 2018 | 5% | 64% | 26% | 4% | 2% |
| 2019 | 5% | 65% | 25% | 3% | 2% |
| 2020 | 4% | 65% | 26% | 3% | 2% |

OC has been an educational institution of choice for students in the neighboring geographic region and for those who are often considered under-represented in higher education and need financial assistance/aid. First generation students make up a sizable portion of OC's student population. As stated in its mission statement, OC emphasizes the use of active learning practices and innovative use of technology to enhance student success and learning. Consequently, a critical component of the technology infrastructure at OC is its LMS, which currently is D2L's online learning platform, Brightspace.

### 1.2 About Brightspace

The LMS systems have gained rapid implementations in academic institutional settings to support fully in-person, online, and hybrid learning paradigms. They have been widely incorporated in K-12 and higher educational institutions to support face-to-face, online and, hybrid teaching modalities [8, 16]. There are a variety of LMSs currently being used by academic institutions, and each has various features and tools. The adoption and functional utilization of LMSs are determined by a variety of factors such as compatibility of the system with the requirements of an academic institution, relative advantages of a new LMS over an existing system, the availability of a system to be pilot-tested, and ease of use [8, 29]. Other LMSs features, such as system's uptime, security, scalability, and functionality have also been considered as crucial factors in determining its success. We believe that, going forward, incorporation and availability of a suite of analytical capabilities incorporated into LMSs may also become a key factor in differentiating LMSs available in the market.

In this study, we used Brightspace as the LMS, which was administered and maintained by OC. Brightspace is an open software platform owned by Desire2Learn (D2L). The privately held, D2L software company was founded in 1999, and its LMS product, Brightspace, has since attained around 12% of the higher education (post K-12) market share according to an analysis [26]. As of 2019, Brightspace was placed among the top four LMSs used in higher education setting by enrollment metric. It has been reported that Brightspace has a higher adoption than Moodle, and a lower adoption than its competing products, Canvas and Blackboard LMSs [18, 26]. Typically, LMSs feature an ecosystem of features such as assignment dropbox, quiz-related tools, content management features, activity calendars, video-conferencing options, discussion forums, and other tools that support the learning experience. In particular, Brightspace features an online learning environment that provides web-based tools for several aspects of teaching and learning process, including hosting lecture presentations, assignment dropboxes, quiz tools, activity calendars, discussion forums, announcements, course content hosting, gradebook, etc. As of 2020, a few newer features of Brightspace include apps for accessing content on mobile devices, analytics of learner performance, audio- and video-based instructor feedback, and course gamification [12].

In 2014, D2L announced that its purpose and vision is to enhance learning through the use of technology [6]. With Brightspace's integration of an adaptive learning solution (D2L LeaP), it provides good analytics, game-based learning, and an expanded mobile-compatible system. On the analytics technology side, IBM's Cognos is embedded into D2L's analytics platform. This is engineered to help faculty drive student learning and to provide an enterprise-class business intelligence tool to the education industry. With IBM, D2L is able to provide the tools to integrate and analyze vast amounts of learner data – and to provide a single, unified view of that data for all decision makers [13].

## 2 RELATED WORK

There have been previous works [14, 15, 23] on measures and metrics used to analyze the usage of LMS data logs to investigate relationships between online learning and student's academic performance.

### 2.1 Analyzing Student Activity and Interaction Patterns in LMS

One of the approaches to detect and quantify academically at-risk students is by using metrics obtained from application of machine learning algorithms to data logs provided by the LMS [24]. In this study, several quantitative measures and data sources were used including student attendance, the number of log-ins, the number of students starting a lesson, number of assigned activity submission completions, and the duration of log-in time. The study used machine learning methods such as logistic regression, support vector machine, and random forest to predict student Grade Point Average (GPA). These authors found the accuracy of the logistic regression model to be relatively better than the other models. However, the random forest model detected more at-risk students in the early stages of the course. These results imply that the usage of LMS data and the implementation of machine learning methods can be used to detect academically at-risk students based on their patterns of interaction, and in particular, can be used for early detection of off-task behavior of students (i.e., behavior not related to course lessons or objectives).

Cerezo *et al.* examined students' asynchronous learning processes via an educational data mining approach using data extracted from a Moodle LMS logs of students, who were grouped according to similar behaviors in terms of effort, time spent working, and procrastination [10]. Methods used in the study included cluster analysis to group students with similar behaviors and ANOVA to observe the hypothesized inter-cluster differences. The authors grouped students into four major clusters and found that i) time, ii) task, and iii) the time it took for students to turn in an assignment (after the assignment was available) were the three variables most related with the students' final grade. These results imply that the more time students dedicate to working on tasks, the less time they take to turn in the assigned tasks. Furthermore, the authors concluded that grouping students in clusters that shared similar behavior was beneficial in determining the significant differences and actions (of students) when conducting multiple comparisons of the variables. By grouping such students, it can help instructors better understand students' learning characteristics and interaction with an LMS. In particular, this method can help to identify academically at-risk students. In the study, the authors categorized such at-risk students as having non-task oriented or procrastinating features.

### 2.2 Predicting Student Performance in LMS

An approach which discovers potential relationships among learners' performance, course characteristics, and the usage of an LMS is proposed [22]. The measures used in this study included, but were not limited to, the total number of sessions per course viewed by all users, the total number of visits per course by all users, and the duration of total visits per course by all users. The data mining techniques applied to the LMS data included classification, clustering, association rule mining, and regression analysis. These authors found that the classification and the clustering mining techniques showed the number of visits and duration in a course in LMS lead to a positive learners' performance. The regression analysis technique showed the most significant attribute correlated to student performance was the duration of study. These results imply that students' behavior in an online learning platform influences student performance and that it is necessary to collect the LMS data to track students' activity to predict student performance.

An examination was conducted on the significant behavioral indicators of learning using LMS data that may predict course achievement [31]. The study included the frequency measures, the duration of study time, and other elaborated time measures that are indicative of self-regulated learning. In addition, the study examined the potential use of early prediction based on student data which was collected in the middle of the course to investigate the following research questions: 1) "Which indicators from LMS data significantly predict online course achievement?"; and 2) "Do significant indicators and performance data collected in the middle of the course significantly predict online course achievement?" To investigate these research questions, descriptive statistics and correlation analyses were performed, and hierarchical regression analyses were conducted to identify significant LMS measures. During the middle of a course, the authors collected and analyzed LMS data on regular study (i.e., elaborated time-based measure), number of sessions, late submissions, and proof of reading course. These data were found to be good predictors of midterm exam scores and significantly determined final course outcome. Furthermore, the authors concluded that there is a benefit to tracking and analyzing LMS data throughout the duration of courses. Using students' performance data during the course may help instructors predict final course achievement and provide meaningful feedback and interventions to students. These results imply that the extraction and aggregation of indicators from students-LMS' interaction data and the development of reliable prediction models are valuable for educators who seek to understand and improve their students' learning statuses.

Conijn *et al.* furthered the analysis of the portability of prediction models and the accuracy of timely prediction [11]. Their analysis was based on a study of 17 blended, undergraduate courses (traditional face-to-face and fully online courses) to predict student performance from LMS predictor variable(s) and from in-between assessment grades. Their study also analyzed whether it was possible to identify students at-risk early-on in a course, and to what extent these data-based models may be used to generate targeted interventions. In addition, an analysis was conducted to determine how the prediction accuracy of student performance improved as the course progressed. The predictor variables included (but were not limited to), total number of clicks, number of online sessions, number of course page views, irregularity of study time, irregularity of study interval, largest period of inactivity (minutes), time until first activity (minutes), number of links viewed, number of quizzes started, and number of attempts per quiz. To measure the portability of the prediction models, multi-variate analyses and ordinary least squares regression were used. Multiple linear regressions and logistic regression were used to determine the predictability of the models for appropriate variables. Pearson correlation analyses were performed to determine the correlation between the predictor variables and the final exam grade. The authors reported that there was no comprehensive set of variables that consistently predicted student performance across multiple courses. Furthermore, the authors concluded that after controlling for a large set of predictor variables, the portability of the prediction models across courses is low, and could be explained by varying instructor-specific learning designs, as the type of the assigned pedagogical activities available in an LMS can influence the number of LMS visits.

In another study, instead of focusing on the exam performance, LMS data was used to predict assignment success from partial data before the completion of the course: at 10%, 25%, 33%, and 50% levels of course completion [28]. Their aim was to detect at-risk, fail, and successful students early in the course. Machine learning methods of decision tree, naive Bayes, logistic regression, multilayer perceptron (MLP) neural network, and support vector machine were used to make the predictions. Accuracies of all the methods increased with the amount of data, as expected. They found that the MLP method showed the best prediction performance overall. The authors also used clustering to divide the students into six groups based on their interaction with the LMS and found strong correlations of four of these groups with the students' academic performance.

## 3 METHODS

### 3.1 Datasets

An Institutional Review Board (IRB) request was submitted to OC to get access to archival data sets of LMS data for all courses offered by OC during fall 2019, spring 2020, and fall 2020 semesters. After the IRB request was approved, a data request was submitted to campus LMS administrator. Two data reports were exported from college-wide LMS:

1) Learner Usage Report (LUR) that documents all of students' activity on LMS and
2) Grades Report (GR) that contains the grades for graded events in all categories.

For LUR, a total of 234,437 entries were returned, and for GR, a total of 3,258,478 entries were returned. To maintain anonymity, each student was assigned a unique ID to replace User_Id and Org_Defined_Id. The unique ID was used to combine LUR and GR. A detailed description of variables used in this paper can be found in Appendix A.

*3.1.1 Course Filtering*

During the period studied (fall 2019, spring 2020, and fall 2020), a total of 7,484 sections of courses were offered across multiple schools (School of Business, School of Education, School of Health and Sciences, School of Liberal Arts, School of Science and Technology, Career Center, and Center for Teaching of Excellence). The courses administered by Career Center and Center for Teaching of Excellence for training sections and workshops were excluded from our analysis because they are not full-semester academic courses. Also, summer semester courses were not included due to a much higher percentage of transient students who normally are enrolled in other institutions during spring and fall semesters.

*3.1.2 Feature Selection*

LUR contains data about interaction between students and LMS. Data from LUR may be categorized into logically connected variables such as:

- System Access (Last_System_Login, Last_Visited_Date, Number_Of_Logins_To_The_System),
- Course Content Access (Content_Completed, Content_Required, Checklist_Completed, Total_Time_Spent_In_Content),
- Quiz Submission and Attempts (Quiz_Completed, Total_Quiz_Attempts),
- Discussion Forum Interaction (Discussion_Post_Created, Discussion_Post_Replies, Discussion_Post_Read, Last_Discussion_Post_Date), and
- Assignment Submission (Number_Of_Assignment_Submissions, Last_Assignment_Submission_Date).

Four features with less than 20% data missing were selected to be used in further analysis:

1) Content_Completed,
2) Number_Of_Assignment_Submissions,
3) Total_Time_Spent_In_Content, and
4) Number_Of_Logins_To_The_System.

To account for the various levels of LMS usage among sections taught by different instructors, each feature was normalized using the maximum and minimum values of each individual section (see equation below). To examine the multicollinearity among these four features, variance inflation factor (VIF) was calculated for each feature. The presence of multicollinearity among features would cause inflated variance resulting in a large VIF [25]. All of the VIFs were less than 10 (Table 4), indicating that no significant multicollinearity was present. Therefore, all four features mentioned above were used in the subsequent analysis.

$$normalized\ value = \frac{x - x_{min}}{x_{max} - x_{min}}$$

Table 4: Variance Inflation Factors for selected features

| Features | VIF |
|---|---|
| Content_Completed | 4.029738 |
| Number_Of_Assignment_Submissions | 3.423563 |
| Total_Time_Spent_In_Content | 2.317332 |
| Number_Of_Logins_To_The_System | 2.493167 |

*3.1.3 Dependent Variable*

Final grades were used as the dependent variable. There are two types of final grades reported in the LMS: final calculated grades and final adjusted grades. Final calculated grades were used as the primary dependent variable, and final adjusted grades were used whenever final calculated grades were unavailable. Since our focus was to identify at-risk students, final grades were converted to a binary variable using 70% as the threshold. Students with final grades less than 70% were considered as at-risk students. At-risk students comprised 18.5% of the dataset. Random oversampling was used to balance the data.

### 3.2 Supervised Learning

The dataset was split into training set (80% of the instances) and testing set (20% of instances). Seven algorithms were used to build binary classifiers: logistic regression, naive Bayes, decision tree, random forest, neural network, XGBoost, and support vector machine. All algorithms were implemented from the scikit-learn 0.24.1 framework. All features were already normalized between 0 to 1 as described in Section 3.1.2. Analysis coding was implemented in Jupyter Notebook 6.0.3. Exhaustive Grid Search using 3-fold cross validation from scikit-learn was used to tune hyper-parameters. Accuracy metric was used to assess the performance of each hyper-parameter combination. A detailed list of all hyper-parameters used for each algorithm can be found in Appendix A Table A1.

SHAP (SHapley Additive exPlanation) is a method for measuring feature influence that was initially proposed in 1953 [30]. It uses game theory to assign a value, SHAP value, to each feature for each observation, which can be used to assess the influence of a specific feature on the final output of the model. SHAP values were used to rank feature importance in all seven algorithms.

### 3.3 Unsupervised Learning

The four features mentioned in section 3.1.2 were also used as input data for unsupervised learning: cluster analysis. K-means was selected as the clustering algorithm [17]. In k-means, all data was partitioned into k randomly generated non-overlapping clusters. Each data point was assigned to a specific cluster so that the squared Euclidean distance between the data point and the cluster center was the smallest. After all data points were assigned, each cluster center was updated by the arithmetic mean of all data points in that cluster. This process was repeated until all cluster centers stopped moving. The effect of this process is to keep similar data points in the same cluster, while maximizing the dissimilarity between the different clusters. In order to determine the optimal number of clusters (*k*) in the k-means algorithm, an elbow method was used [7]. Specifically, the sum of squared Euclidean distances between all data points and corresponding cluster centers was plotted as a function of *k* (Figure 1A). As *k* increased, the sum of the squared

Euclidean distances decreased, with a constantly increasing slope. The value of *k* at which the slope started to level off was identified as the elbow, and the corresponding *k* value, 4, was established as the number of clusters. Silhouette coefficient was also used to determine the optimal number of clusters [5]. Silhouette coefficient is a measure of how similar a data point is to its own cluster comparing to all data points in other clusters. Mean of Silhouette coefficient for each data point was plotted as a function of *k* (Figure 1B). As *k* increased, mean Silhouette coefficient was found to decrease. The optimal number of clusters was determined to be 2, which has the highest mean Silhouette coefficient. Both 2 and 4 clusters were used in the subsequent cluster analysis. The K-means algorithm implementation from the scikit-learn 0.24.1 was used for cluster analysis.

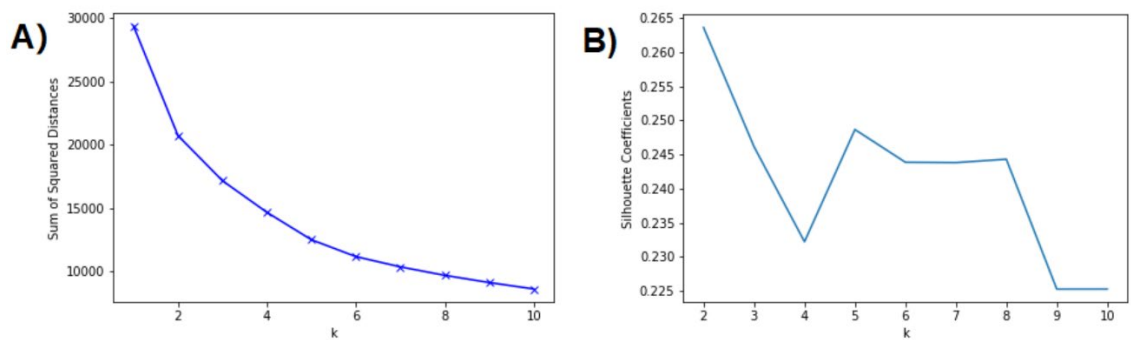

Figure 1: Determination of optimal number of clusters using A) elbow method and B) Silhouette coefficients.

Analysis of variance (ANOVA) [27] was used to test if mean values of all four features and final grades were significantly different among clusters. Tukey's Honest Significant Difference (HSD) post-hoc tests [1] that compares all possible pairs of means was also used to identify which specific clusters' feature/final grades were different.

### 3.4 Supervised Learning

Accuracies and Area Under the Curve (AUC) for all seven algorithms implemented in this work are summarized in Table 5. AUC measures the entire two-dimensional area underneath the Receiver Operating Characteristic (ROC) curve. Algorithm with the best performance is labeled in bold red in Table 5.

Table 5: Accuracies and AUC for seven algorithms

| Algorithm | Accuracy (%) | AUC |
| --- | --- | --- |
| Logistic Regression | 72.16 | 0.79309 |
| Naive Bayes | 71.14 | 0.78447 |
| Decision Tree | 91.14 | 0.90977 |
| **Random Forest** | **94.75** | **0.98697** |
| Neural Network | 72.83 | 0.80085 |
| XGBoost | 93.34 | 0.97183 |
| Support Vector Machine | 73.04 | 0.73033 |

Among all seven algorithms, random forest outperformed the rest algorithms with the highest accuracy (94.75%) and the largest AUC (0.98697). Other tree-based algorithms, decision tree and XGBoost, also performed well with accuracies above 90% and AUC values above 0.9.

## 3.5 Feature Importance

In order to interpret relative feature importance for machine learning models, SHAP values were calculated for all seven algorithms (Figure 2). Each point on the Figure 2 is a Shapley value for a feature and an instance. The position on the y-axis is determined by the feature importance and on the x-axis by the Shapley value. Features are ordered in the descending importance from top to bottom. Relative feature importance for all seven algorithms are summarized in Table 6.

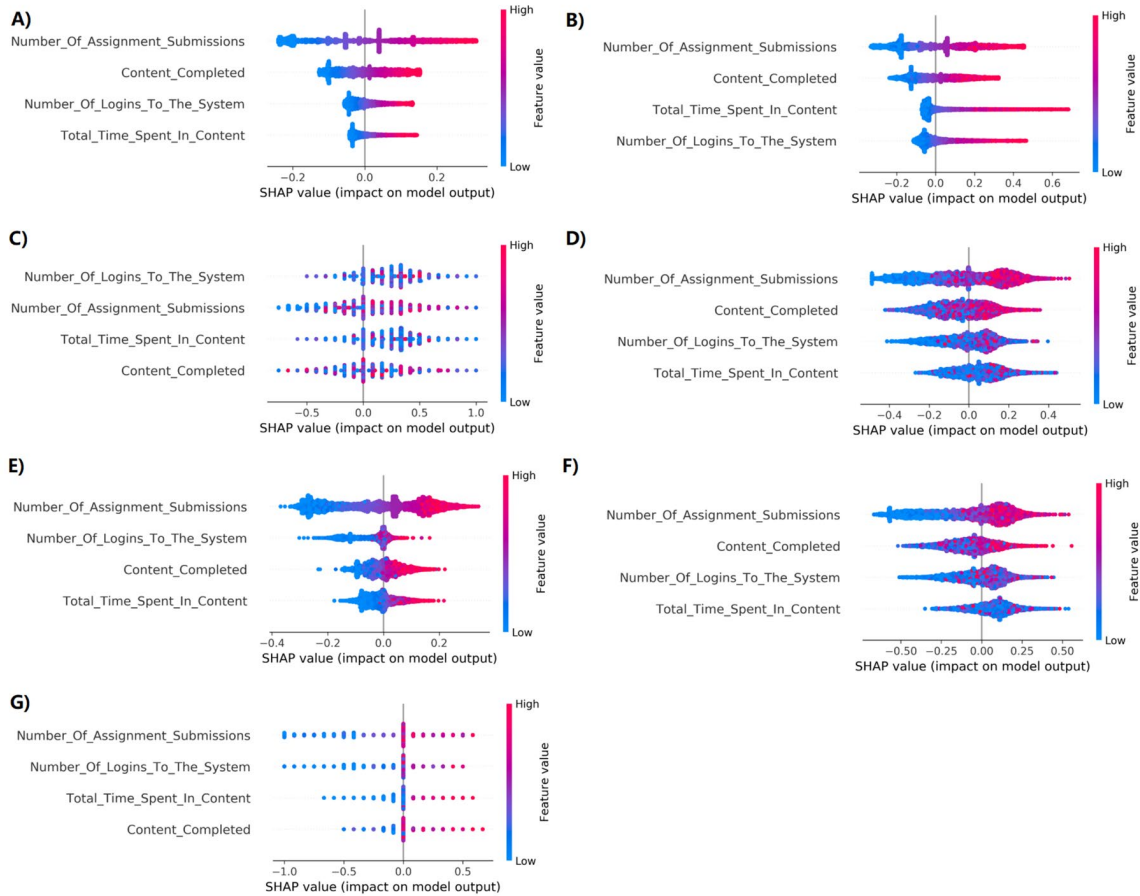

Figure 2: SHAP summary plots to rank feature importance for A) logistic regression, B) naive Bayes, C) decision tree, D) random forest, E) neural network, F) XGBoost, and G) support vector machine.

Table 6: Feature importance rank score for seven algorithms (smaller indicates more feature importance).

| Algorithm | Number_Of_Assignment_Submissions | Content_Completed | Total_Time_Spent_In_Content | Number_Of_Logins_To_The_System |
|---|---|---|---|---|
| Logistic Regression | 1 | 2 | 4 | 3 |
| Naïve Bayes | 1 | 2 | 3 | 4 |
| Decision Tree | 1 | 4 | 3 | 2 |
| Random Forest | 1 | 2 | 4 | 3 |
| Neural Network | 1 | 3 | 4 | 2 |
| XGBoost | 1 | 2 | 4 | 3 |
| Support Vector Machine | 1 | 4 | 3 | 2 |
| Overall | 1 | 2 | 4 | 2 |

Among all seven algorithms studied, **Number_Of_Assignment_Submissions** feature was always ranked as the most important feature followed by two features in tie with each other: **Content_Completed** and **Number_Of_Logins_To_The_System**. The feature, **Total_Time_Spent_In_Content** was ranked as the least important feature considering all seven algorithms.

### 3.6 Clustering

As described in Section 3.3, the optimal number of clusters was determined to be four using elbow method, whereas two clusters were determined as optimal using Silhouette coefficients. In this section, clustering algorithms and analysis results using both two clusters and four clusters are discussed.

#### 3.6.1 Four Cluster Analysis

Clustering analysis of students' interaction with LMS are summarized in Figure 3. Figure 3 shows the mean and standard deviation values of each cluster mentioned in Table 7. Since our goal was to identify if any correlation existed between each cluster and corresponding end of semester grades, mean and standard deviation of final grades are also shown in Figure 3 and Table 7.

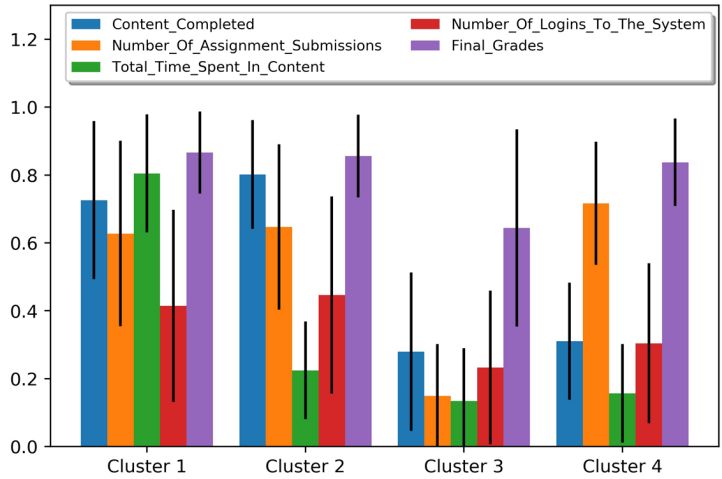

Figure 3: Bar plot of centroids using four clusters.

Table 7: Number of students, mean ± standard deviation values for each cluster using four clusters.

| Cluster# | Count of Students | Content_Completed | Number_Of_Assignment_Submissions | Total_Time_Spent_In_Content | Number_Of_Logins_To_The_System | Final_Grades |
|---|---|---|---|---|---|---|
| Cluster-1 | 13756 | 0.726 ± 0.233 | 0.627 ± 0.273 | 0.805 ± 0.174 | 0.414 ± 0.283 | 0.866 ± 0.121 |
| Cluster-2 | 24942 | 0.802 ± 0.160 | 0.647 ± 0.244 | 0.224 ± 0.144 | 0.446 ± 0.290 | 0.856 ± 0.122 |
| Cluster-3 | 22646 | 0.279 ± 0.233 | 0.150 ± 0.152 | 0.134 ± 0.155 | 0.233 ± 0.226 | 0.644 ± 0.290 |
| Cluster-4 | 22662 | 0.310 ± 0.173 | 0.716 ± 0.182 | 0.157 ± 0.145 | 0.304 ± 0.235 | 0.837 ± 0.129 |

Cluster-1 contained the fewest number of students with the highest average final grades. Also, students in Cluster-1 interacted with LMS more actively (higher average values) in all of the four features. Students in Cluster-2 had slightly lower average final grades. They interacted with LMS even more actively in all of the features except for a much lower value of *Total_Time_Spent_in_Content*. Students in Cluster-3 had the lowest values for every feature as well as the final grades among all four clusters, which indicated lack of engagement throughout the semester. Lastly, students in Cluster-4 had slightly lower final grades compared to Cluster-1 and Cluster-2. However, they had a much lower values for *Content_Completed* and *Total_Time_Spent_In_Content*.

According to ANOVA, significant difference was observed only for **Number_Of_Assignment_Submissions** among different clusters ($p < 0.05$). **Content_Completed** was marginal statistically different ($p = 0.0829$). Tukey's post-hoc test was used to reveal that between Cluster-2 and Cluster-3, mean values of **Number_Of_Assignment_Submissions** were statistically different from each other at a significance level of 0.05.

### 3.6.2 Two Cluster Analysis

Clustering analysis was repeated using two clusters. Clustering results including mean and standard deviation values of each cluster were shown in Figure 4 and summarized Table 8. Note that final grades were not used as input for clustering, they are included in Figure 4 and Table 8 to assist with interpreting clustering results.

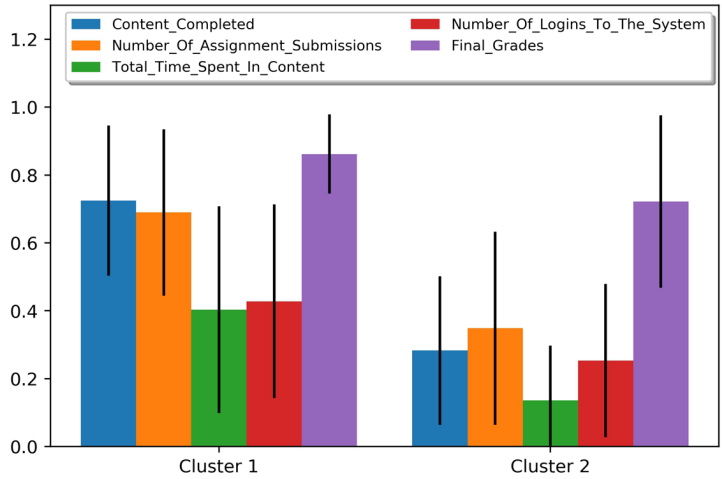

Figure 4: Bar plot of centroids using two clusters.

Table 8: Number of students, mean ± standard deviation values for each cluster using two clusters.

| Cluster# | Count of Students | Content_Completed | Number_Of_Assignment_Submissions | Total_Time_Spent_In_Content | Number_Of_Logins_To_The_System | Final_Grades |
| --- | --- | --- | --- | --- | --- | --- |
| Cluster-1 | 44318 | 0.725 ± 0.221 | 0.690 ± 0.245 | 0.403 ± 0.305 | 0.428 ± 0.285 | 0.862 ± 0.117 |
| Cluster-2 | 39688 | 0.282 ± 0.219 | 0.348 ± 0.284 | 0.136 ± 0.161 | 0.253 ± 0.226 | 0.722 ± 0.253 |

Both clusters had a similar/comparable number of students. Students in Cluster-1 had much higher values in all features as well as final grades. On the other hand, students in Cluster-2 were not as involved in interaction with LMS. According to ANOVA, significant differences were observed for **Number_Of_Assignment_Submissions** and **Content_Completed** among different clusters ($p < 0.05$).

## 4 DISCUSSION AND FUTURE WORK

In this work, we demonstrated the use of data analytics using the data logs provided by Brightspace LMS to promote a better learning and teaching experience at a 4-year undergraduate degree-awarding college. As discussed, in this study, LMS data may be used to identify academically at-risk students during an ongoing semester. We believe that having a LMS data-driven mechanism to identify at-risk students is essential for educators and administrators to plan and implement appropriate academic interventions during the semester. The current study used LMS data from fall 2019, spring 2020, and fall 2020 semesters to identify observable features that can be used to predict whether a student would pass/fail a course. This is a first step towards the goal of identifying at-risk students during the progression of a semester, rather than at the end. This ability afforded by the LMS data allows students and instructors to implement interventions during the semester to support student success. In this study, we implemented and compared the accuracies of seven supervised learning algorithms in identifying at-risk students. We found that the random forest algorithm provided the highest prediction accuracy. The other tree-based algorithms, decision tree and XGBoost, were also found to perform well and produced > 90% prediction accuracy. From our study based on historical full-semester data, we stipulate that tree-based algorithms are suitable for LMS data and we believe that they may also perform well in situations using partial/early semester data to predict student performance.

From the available data logs captured by Brightspace, we found the feature, ***Number_Of_Assignment_Submissions***, to be ranked as the most important feature by all seven supervised learning algorithms mentioned in Table 6. The feature, ***Content_Completed***, was found to be ranked as the second most important feature by four (logistic regression, naive bayes, random forest, and XGBoost) of the seven implemented algorithms. These observations are consistent with and in alignment with instructors' observations of students not completing assignments and not viewing the course materials posted in the content area of the LMS. In our study, we found these two factors to be good indications of eventual failure in a course.

In addition to the seven supervised learning algorithms, we also used unsupervised learning algorithm k-means to group students into different clusters in terms of their similarities in their features recorded by the LMS data. Features are observable students' interaction documented in the LMS, and we are interested in their relevance to the final grade. In this study, we chose four Brighspace's features (***Content_Completed, Number_Of_Assignment_Submissions, Total_Time_Spent_In_Content, Number_Of_Logins_To_The_System***) for both supervised learning and unsupervised learning.

The two-cluster study was most revealing about the feature importance. The Cluster-1 of students were found to be more active learners (engaged in LMS) than those in the Cluster-2. The two features in which Cluster-1 and Cluster-2 are statistically different are ***Number_Of_Assignment_Submissions*** and ***Content_Completed.*** Noticeable, these two features were also identified as the two most important in predicting the final grade by the supervised learning algorithm. Not surprisingly, Cluster-1 students had much higher final grades than Cluster-2 students did.

The four-cluster study also reinforced the above observations of feature importance. The features that were significantly different among the four clusters were ***Number_Of_Assignment_Submissions*** and ***Content_Completed*** (albeit marginally), the same as those identified in the supervised learning and the two-cluster study. The most distinct cluster is Cluster-3, which has the lowest values for every feature, which indicates lack of engagement throughout the semester. Not surprisingly, Cluster-3 also has the lowest final grades. The other three clusters (Cluster-1, Cluster-2, and Cluster-4) show high values along all or most of the features. However, we observed a decreasing student-engagement (with LMS) from Cluster-1 to Cluster-2 and then again to Cluster-4. The Cluster-1 had high values for all features, indicating high student engagement with LMS. The Cluster-2 was similar to Cluster-1, except for a much lower value of ***Total_Time_Spent_in_Content***. The Cluster-4 was similar to Cluster-2, except for a much lower value of ***Content_Completed***. The Cluster-2's final grade was only slightly lower than that for Cluster-1, from which we concluded that the impact of the ***Total_Time_Spent_in_Content*** feature was not significant in the final grade. The Cluster-4's final grade's drop from Cluster-2 was larger than the drop from Cluster-1 to Cluster-2, which indicated that the impact of the ***Content_Completed*** feature on the final grade was more significant.

In our experience, the ***Total_Time_Spent_in_Content*** feature was not that important because many instructors posted links to external resources in the LMS. The students downloaded the external resources whose link was posted in the LMS and reviewed them independently, without using LMS. As a result, the ***Total_Time_Spent_in_Content*** feature in the LMS may undercount the amount of time that students actually spent in studying.

To summarize, both the 2-cluster and the 4-cluster unsupervised learning algorithms identified the same set of two features that were statistically different among clusters, they are ***Number_Of_Assignment_Submissions*** and ***Content_Completed***. Multiple supervised learning algorithm also identified the same set of two features that were most important for predicting students' final grades. The arrival at the same conclusion from two different and independent approaches (supervised and unsupervised algorithms) is a highlight of our study.

As an extension to this study, we plan to compute metrics/features that may be engineered using the early semester LMS data and to identify academically at-risk students. We believe that identifying at-risk students during early semester provides an opportunity to educators and students to realign their academic efforts to promote success in a course of study. We are also interested in investigating the use of just two features (**Number_Of_Assignment_Submissions** and **Content_Completed**) to predict a student's final performance at different points of time during a semester.

Lastly, the current study lays the groundwork and provides a proof-of-concept for building a real-time data analytics/mining tool that may be developed and incorporated inside a deployed LMS. Other suggested extensions of our study include the changes of student behavior due to the COVID-19 pandemic. A quick study of the LMS data from fall 2019 to fall 2020 revealed that a noticeable increase in LMS usage, which may be attributed to the onset of COVID-19 pandemic. In future study, we plan to understand the systematic pattern of changes in LMS usage, specifically, how student engagement has evolved from the start of pandemic, during pandemic before vaccines were available, and when schools opened up after vaccines became available. Additionally, we are also interested in studying whether an increased LMS usage may make predictions more/less accurate and what impact they may have on students' academic performance.

## A    APPENDIX

### A.1    Description of variables used in the analysis

1. Org_Defined_Id: Unique ID created by LMS for every user
2. Role_Name: denotes students, TAs, or instructors
3. Content_Completed: Number of content items completed by a student for a certain course
4. Content_Required: Number of content items required by instructors for a certain course
5. Checklist_Completed: Number of checklists completed by a student for a certain course
6. Quiz_Completed: Number of quizzes completed by a student for a certain course
7. Total_Quiz_Attempts: Total number of attempts to quizzes by a student for a certain course
8. Discussion_Post_Created: Number of posts created by a student in discussion forum for a certain course
9. Discussion_Post_Replies: Number of posts replied by a student in discussion forum for a certain course
10. Discussion_Post_Read: Number of posts read by a student in discussion forum for a certain course
11. Last_Discussion_Post_Date: Date of the last post on discussion forum
12. Number_Of_Assignment_Submissions: Total number of assignments submitted on LMS for a specific course
13. Last_Assignment_Submission_Date: Date of submitting the last assignment

14. Total_Time_Spent_In_Content: Total time spent in content on LMS (min) for a specific course
15. Last_Visited_Date: Date of last visit to specific course
16. Last_System_Login: Date of last login to LMS
17. Number_Of_Logins_To_The_System: Total number of logins to LMS
18. Grade_Item_Name: Denotes final calculated grades, final adjusted grades, and grades for any other graded events on LMS
19. Grade_Value: Value of grades

## A.2  Hyper-parameters for final optimized models

Table A1. Optimized hyper-parameters used in supervised learning algorithms.

| Algorithm | Optimized Hyperparameters |
|---|---|
| Logistic Regression | 'C': 0.009, 'class_weight': None, 'dual': False, 'fit_intercept': True, 'intercept_scaling': 1, 'l1_ratio': 0, 'max_iter': 100, 'multi_class': 'auto', 'n_jobs': None, 'penalty': 'l2', 'random_state': None, 'solver': 'liblinear', 'tol': 0.0001, 'verbose': 0, 'warm_start': False |
| Naive Bayes | 'priors': None, 'var_smoothing': 1e-09 |
| Decision Tree | 'ccp_alpha': 0.0, 'class_weight': None, 'criterion': 'gini', 'max_depth': None, 'max_features': None, 'max_leaf_nodes': None, 'min_impurity_decrease': 0.0, 'min_impurity_split': None, 'min_samples_leaf': 1, 'min_samples_split': 2, 'min_weight_fraction_leaf': 0.0, 'random_state': None, 'splitter': 'best' |
| Random Forest | 'bootstrap': True, 'ccp_alpha': 0.0, 'class_weight': None, 'criterion': 'gini', 'max_depth': None, 'max_features': 'log2', 'max_leaf_nodes': None, 'max_samples': None, 'min_impurity_decrease': 0.0, 'min_impurity_split': None, 'min_samples_leaf': 1, 'min_samples_split': 3, 'min_weight_fraction_leaf': 0.0, 'n_estimators': 700, 'n_jobs': None, 'oob_score': False, 'random_state': None, 'verbose': 0, 'warm_start': False |
| Neural Network | 'activation': 'tanh', 'alpha': 0.1, 'batch_size': 'auto', 'beta_1': 0.9, 'beta_2': 0.999, 'early_stopping': False, 'epsilon': 1e-08, 'hidden_layer_sizes': (100, 100), 'learning_rate': 'adaptive', 'learning_rate_init': 0.001, 'max_fun': 15000, 'max_iter': 1000, 'momentum': 0.9, 'n_iter_no_change': 10, 'nesterovs_momentum': True, 'power_t': 0.5, 'random_state': None, 'shuffle': True, 'solver': 'lbfgs', 'tol': 0.0001, 'validation_fraction': 0.1, 'verbose': False, 'warm_start': False |
| XGBoost | 'objective': 'binary:logistic', 'use_label_encoder': True, 'base_score': 0.5, 'booster': 'gbtree', 'colsample_bylevel': 1, 'colsample_bynode': 1, 'colsample_bytree': 1, 'gamma': 0, 'gpu_id': -1, 'importance_type': 'gain', 'interaction_constraints': '', 'learning_rate': 0.1, 'max_delta_step': 0, 'max_depth': 15, 'min_child_weight': 1, 'missing': nan, 'monotone_constraints': '()', 'n_estimators': 400, 'n_jobs': 4, 'num_parallel_tree': 1, 'random_state': 0, 'reg_alpha': 0, 'reg_lambda': 1, 'scale_pos_weight': 1, 'subsample': 1, 'tree_method': 'exact', 'validate_parameters': 1, 'verbosity': None |
| Support Vector Machine | 'C': 1000, 'break_ties': False, 'cache_size': 200, 'class_weight': None, 'coef0': 0.0, 'decision_function_shape': 'ovr', 'degree': 3, 'gamma': 1, 'kernel': 'rbf', 'max_iter': -1, 'probability': False, 'random_state': None, 'shrinking': True, 'tol': 0.001, 'verbose': False |